\PassOptionsToPackage{pdfpagelabels=false}{hyperref}   % Fix warning pdfpagelabels is turned off
\expandafter\def\csname ver@fixltx2e.sty\endcsname{}   % Fix warning fixltx2e is not required

\documentclass[fleqn,usenatbib]{mnras}

\usepackage{graphicx}	% Including figure files
\usepackage{amsmath}	% Advanced maths commands
\usepackage{amssymb}	% Extra maths symbols
\usepackage{siunitx}
\usepackage{makecell}

\usepackage{txfonts}

\usepackage[T1]{fontenc}
\usepackage{ae,aecompl}

\title[The G-dwarf problem revisited]{SDSS-IV MaNGA: the ``G-dwarf problem'' revisited}

\author[M. J. Greener et al.]{
Michael J. Greener,$^{1}$\thanks{E-mail: michael.greener@nottingham.ac.uk}
Michael Merrifield,$^{1}$
Alfonso Arag{\'o}n-Salamanca,$^{1}$
Thomas Peterken,$^{1}$
\newauthor
Brett Andrews,$^{2}$
and Richard R. Lane$^{3}$
\\
$^{1}$School of Physics \& Astronomy, University of Nottingham, University Park, Nottingham, NG7 2RD, UK\\
$^{2}$Department of Physics and Astronomy, University of Pittsburgh, 3941 O'Hara Street, Pittsburgh, Pennsylvania 15260, USA\\
$^{3}$Instituto de Astronom{\'i}a y Ciencias Planetarias de Atacama, Universidad de Atacama, Copayapu 485, Copiap{\'o}, Chile\\
}

\date{Accepted XXX. Received YYY; in original form ZZZ}

\pubyear{2020}

\begin{document}
\label{firstpage}
\pagerange{\pageref{firstpage}--\pageref{lastpage}}
\maketitle

\begin{abstract}
The levels of heavy elements in stars are the product of enhancement by previous stellar generations, and the distribution of this metallicity among the population contains clues to the process by which a galaxy formed. Most famously, the ``G-dwarf problem'' highlighted the small number of low-metallicity G-dwarf stars in the Milky Way, which is inconsistent with the simplest picture of a galaxy formed from a ``closed box'' of gas. It can be resolved by treating the Galaxy as an open system that accretes gas throughout its life. This observation has classically only been made in the Milky Way, but the availability of high-quality spectral data from SDSS-IV MaNGA and the development of new analysis techniques mean that we can now make equivalent measurements for a large sample of spiral galaxies. Our analysis shows that high-mass spirals generically show a similar deficit of low-metallicity stars, implying that the Milky Way's history of gas accretion is common. By contrast, low-mass spirals show little sign of a G-dwarf problem, presenting the metallicity distribution that would be expected if such systems evolved as pretty much closed boxes. This distinction can be understood from the differing timescales for star formation in galaxies of differing masses.

\end{abstract}


%%%%%%%%%%%%%%%%%%%%%%%%%%%%%%%%%%%%%%%%%%%%%%%%%%

%%%%%%%%%%%%%%%%% BODY OF PAPER %%%%%%%%%%%%%%%%%%

\section{Introduction}
\label{sec:Introduction}

Almost all of the elements heavier than helium that we find in our galaxy's stars, the ``metals'', are there because these objects incorporate matter recycled from previous stellar generations, with stars born early on containing less of this enhanced material \citep{Schmidt1963TheMass., Talbot1971TheModel, Tinsley1980EvolutionGalaxies}. There are thus clues to the star-formation history of the Galaxy encoded in the distribution of the metallicity that we find in its stars \citep{Talbot1971TheModel}. This phenomenon can be most simply quantified by the cumulative metallicity distribution function (CMDF), which is just the total mass in stars in which the heavy element fraction is less than $Z$, $M_*(<Z)$.

\begin{figure}
	\includegraphics[width=0.47\textwidth]{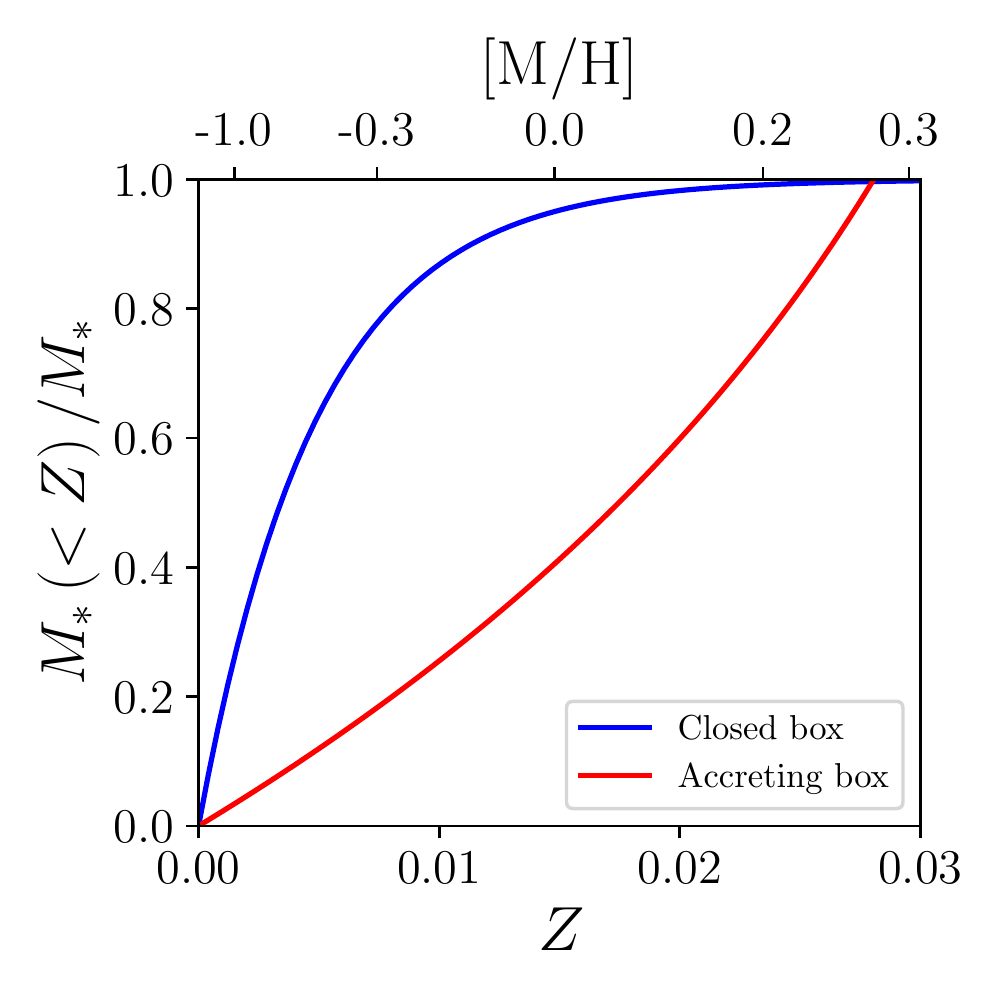}
    \caption{Simple model CMDFs showing the fractional mass of stars that have a metallicity less than $Z$ for a closed box (blue) and an accreting box (red). Characteristically, the yield $p$ of a generation of star formation is of order the value of Solar metallicity ($Z_{\odot} = 0.0148$; \citealp{Lodders2019SolarAbundances}); for these models, we adopt yields of one-third $Z_{\odot}$ for the closed box, and three times $Z_{\odot}$ for the accreting box. These yield values are not physically motivated, but have been selected simply for illustrative purposes.}
    \label{fig:Theoretical_MDFs}
\end{figure}

Such a simple distribution clearly does not contain the full life history of the Galaxy's star formation and gas recycling, but it is sufficiently robust to make quite strong statements about its past history. For example, if the Milky Way formed in isolation from a single initial gas cloud of mass $ M_{\rm{gas,} \: 0}$, with enhanced material well mixed in as it is recycled\footnote{Throughout this work, we adopt the instantaneous recycling approximation, which assumes that metals are expelled by a generation of stars immediately after these stars form (see \citealp{Binney1998GalacticAstronomy} Section~5.3.1).} (a scenario termed the ``closed box'' model of chemical evolution; \citealp{Talbot1971TheModel}; \citealp{Tinsley1974ConstraintsNeighborhood}), then the CMDF takes the simple form 
\begin{equation}
    M_{*} \left( < Z \right) = M_{\rm{gas,} \: 0} \left[ 1 - \exp{ \left( -Z / p \right)} \right],
	\label{eq:closed_box}
\end{equation}
where $p$ is a parameter that defines the yield of heavy elements created by each generation of stars (see, for example, \citealp{Binney1998GalacticAstronomy} Section~5.3.1). An illustration of the resulting function is shown in Figure~\ref{fig:Theoretical_MDFs}. A conflict between this model and observation was first noted by \citet{vandenBergh1962TheAbundances.}, who pointed out that the Milky Way contains many fewer low-metallicity G-dwarf stars than the steep initial rise in this function predicts. This ``G-dwarf problem'' has subsequently been observed in populations of K dwarfs \citep{Casuso2004TheDisc} and M dwarfs \citep{Mould1978InfraredDwarfs, Woolf2012TheGalaxy, Woolf2020TheStars}, and seen both in the Solar neighbourhood \citep[e.g.][]{Rocha-Pinto1996TheNeighbourhood, Gratton1996TheNeighbourhood, Chiappini1996TheModel, Holmberg2007TheDisk} and throughout the Galaxy \citep[e.g.][]{Chiappini2001AbundanceWay, Hayden2015CHEMICALDISK}, so is clearly a substantive issue.

In essence, the problem is that by the time a closed box has built up sufficient heavy elements to make stars with high metallicity, there is very little gas left to make new stars, so it will always produce the majority of its stars at low metallicities. A variety of mechanisms have been invoked to seek to resolve the G-dwarf problem (for an extensive list of proposed solutions to the problem, see \citealp{Pagel2009NucleosynthesisGalaxies} Section~8.4). However, conceptually the simplest solution -- and the most widely accepted -- is to introduce a steady stream of pristine gas to the galaxy, the ``accreting box'' model \citep{Tinsley1974ConstraintsNeighborhood, Tinsley1980EvolutionGalaxies}. In this case, the CMDF can be shown to be 
\begin{equation}
    M_{*} \left( < Z \right) = -M_{\rm{gas}} \left[ \ln{ \left(1 - Z / p \right)} \right],
	\label{eq:accreting_box}
\end{equation}
where $M_{\rm{gas}}$ is a constant (\citealp{Binney1998GalacticAstronomy} Section~5.3.3). As can be seen from Figure~\ref{fig:Theoretical_MDFs}, the constant addition of new gas provides the raw material necessary for more star formation at later times, tipping the balance in favour of high-metallicity stars. The resulting change in the shape of the CMDF has been found to largely eliminate the G-dwarf problem both in the Solar neighbourhood \citep{Gratton1996TheNeighbourhood, Chiappini1996TheModel} and across the entire Galaxy \citep{Chiappini2001AbundanceWay, Hayden2015CHEMICALDISK}.

While such a scenario is reassuring for our understanding of the Milky Way, we lack the context to know where our galaxy fits into the wider picture of chemical enrichment. Although metallicity distribution functions can be produced from analysis of resolved stellar populations in Local Group galaxies \citep[e.g.][]{Escala2018ModellingDiffusion, Manning2018FromGiants, Gilbert2019ElementalStream}, for more distant unresolved galaxies all that we know for sure is that the average stellar metallicities of less massive galaxies are lower \citep[e.g.][]{Gallazzi2005TheUniverse, Panter2008TheRecord}. It therefore remains unclear where the Milky Way lies relative to its spiral galaxy peers in terms of its CMDF.

Fortunately, as recent work by \citet{Mejia-Narvaez2020TheSurvey} indicates, the wealth of data obtained by integral field unit (IFU) surveys in the past few years means that we are now in a position to address this question. Observations from the Mapping Nearby Galaxies at Apache Point Observatory (MaNGA) project \citep{Bundy2015OVERVIEWOBSERVATORY} have provided spectra right across the faces of thousands of nearby galaxies. Spectral synthesis fitting with codes such as \texttt{STARLIGHT} \citep{CidFernandes2005Semi-empiricalMethod} can then be used to decompose such spectra into their component stellar populations of differing ages and metallicities. By integrating across all ages and co-adding all the spatial data for each galaxy, we can reconstruct the CMDFs of these spiral systems for comparison with the Milky Way. Clearly, collapsing all this data into a single one-dimensional function is not making full use of all of the information that it contains, but it does offer a simple robust metric of the global metal content of a spiral galaxy. While the quality of the reconstructed CMDFs may not be as high as for our own galaxy, it should be more than adequate to distinguish between the very different functions of Figure~\ref{fig:Theoretical_MDFs}, providing an overview of the metallicity evolution of a complete sample of spiral galaxies in the local Universe.

\section{Data and Analysis}
\label{sec:Observations}

\subsection{The MaNGA Survey}
\label{subsec:MaNGA}

MaNGA \citep{Bundy2015OVERVIEWOBSERVATORY} is part of the fourth generation of the Sloan Digital Sky Survey \citep[SDSS-IV;][]{Blanton2017SloanUniverse}, and has recently completed its mission to acquire spectroscopic observations for 10000 nearby galaxies \citep{Yan2016SDSS-IVQuality, Wake2017TheConsiderations}. The MaNGA survey thus represents a complete sample of these systems in the local Universe. Using hexagonal IFU fibre bundles \citep{Law2015OBSERVINGSURVEY} to feed into a spectrograph \citep{Smee2013THESURVEY, Drory2015THETELESCOPE} mounted on the $2.5\,{\rm m}$ telescope at Apache Point Observatory \citep{Gunn2006TheSurvey}, spectra were obtained across the face of each galaxy out to at least 1.5 effective radii, capturing most of the light from each system. The raw data were reduced and calibrated \citep{Yan2016SDSS-IV/MaNGA:TECHNIQUE} by the Data Reduction Pipeline \citep[DRP;][]{Law2016TheSurvey}, before being processed through the Data Analysis Pipeline \citep[DAP;][]{Westfall2019TheOverview, Belfiore2019TheModeling} to create the data products employed here.

\subsection{Sample Selection}
\label{subsec:Sample Selection}

Since the intent of this paper is to place the Milky Way metallicity data in context, we need to select a sample of comparable spiral galaxies from the full MaNGA data set. Fortunately, the citizen science project Galaxy Zoo 2 \citep[GZ2;][]{Willett2013GalaxySurvey} provides robust classifications of galaxies upon which we can draw. The process that we follow is essentially identical to that described in \citet{Peterken2020SDSS-IVGalaxies}, except that we make use of the more current ninth MaNGA Product Launch (MPL-9) data. The reasoning behind the method adopted here is described in more detail by \citet{Willett2013GalaxySurvey} and \citet{Hart2016GalaxyBias}.

GZ2 classifications are available for a total of 7330 MPL-9 galaxies. From this sample, we first reject 58 galaxies which were flagged by GZ2 as obscured by a star or other artifact. We then ensure each galaxy has a spiral morphology: following the recommendations of \citet{Willett2013GalaxySurvey} we require that $> 43\%$ of $N \geq 20$ respondents observed either spiral features or a disk in the galaxy. This requirement reduces the sample to 5255 potentially spiral galaxies. Since we are seeking a clean sample of spiral systems, we retain only those which are oriented reasonably face-on so that their spiral structure is apparent. Again following \citet{Willett2013GalaxySurvey}, we require that $> 80\%$ of $N \geq 20$ respondents determine that each galaxy is not edge-on, and we also implement a cut based on the photometric axis ratios of the galaxies such that $\frac{b}{a} \geq 0.5$, which is equivalent to an inclination of $i \geq \ang{60}$. This constraint is slightly more stringent than that suggested by \citet{Hart2017GalaxyAngles}, as discussed by \citet{Peterken2020SDSS-IVGalaxies}, and leaves a sample of 1641 reasonably face-on spiral galaxies. Finally, we remove a further 166 galaxies that were flagged for poor data quality by the DRP or had for any reason failed to produce the necessary DAP data sets. Collectively, these criteria produce the final clean sample of 1475 face-on spiral galaxies that are analysed in this work. The galaxies in this final sample have a median redshift of $z = 0.037$. We also note that none of the results depend at all sensitively on the exact sample selection criteria.

\subsection{Spectral Fitting}
\label{subsec:STARLIGHT}

The stellar evolution histories of the sample galaxies were determined using the full-spectrum stellar population fitting code \texttt{STARLIGHT} \citep{CidFernandes2005Semi-empiricalMethod}. \texttt{STARLIGHT} essentially derives a best fit to each spectrum by combining a set of templates of differing ages and metallicities; the process is very similar to that employed by \citet{Greener2020SDSS-IVGalaxies}, and is explained in detail by \citet{Peterken2020SDSS-IVGalaxies}. Here, we summarise the main steps relevant to this work.

After removing any emission lines using the MaNGA DAP and shifting to zero redshift, each spectrum is fitted using a linear combination of the single stellar population (SSP) E-MILES templates of \citet{Vazdekis2016UV-extendedGalaxies}. The E-MILES library of SSP templates is based on the earlier MILES library \citep{Vazdekis2010EvolutionarySystem}, and we adopt a \citet{Chabrier2003GalacticFunction} initial mass function (IMF), the ``Padova'' isochrones of \citet{Girardi1999Evolutionary0.03}, and an appropriately metallicity scaled value for alpha-element enrichment. The E-MILES templates incorporate nine ages $(\log(\rm age / yr) = 7.85, \allowbreak \: 8.15, \: 8.45, \: 8.75, \: 9.05, \: 9.35, \: 9.65, \: 9.95, \: 10.25)$ and six metallicities $([\rm M / H] = -1.71, \: -1.31, \: -0.71, \: -0.40, \: +0.00, \: +0.22)$. Template logarithmic values, $\rm [M / H]$, are then converted to metallicity $Z = Z_{\odot} \times 10^{\rm [M / H]}$. To reproduce younger stellar populations, we include an additional six ages $(\log(\rm age / yr) = 6.8, \: 6.9, \: 7.0, \allowbreak \: 7.2, \: 7.4, \: 7.6)$ and two metallicities $([\rm M / H] = -0.41, \: +0.00)$ from the templates of \citet{Asad2017YoungCMDs}. Apart from adopting the slightly different \citet{Bertelli1994TheoreticalOpacities.} isochrones, these younger templates were generated using exactly the same method as the E-MILES templates. We use the \texttt{STARLIGHT} configuration settings which prioritise robustness over computation times, following the recommendations of \citet{Ge2018RecoveringUncertainties} and \citet{CidFernandes2018OnAlgorithms}, and as fully described and tested by \citet[including Appendix~A]{Peterken2020SDSS-IVGalaxies}.

The result of this fitting process for every spaxel across the face of a spiral galaxy is a set of weights for the mass contribution made by each SSP to the light seen in that spectrum. Co-adding the results from each spaxel then gives a fit to the integrated light from the entire galaxy, with contributions from SSPs spanning the two-dimensional parameter space of metallicity and age. Adding the contributions from SSPs of different ages reduces the data to a one-dimensional function of the contribution from stars of different metallicities to the total mass of that galaxy. Finally, adding together all the contributions from templates with metallicities less than $Z$ produces the required CMDF for the galaxy, $M_{*}(< Z)$.

\section{Results and Discussion}
\label{sec:Chemical Evolution History}

\begin{figure}
	\includegraphics[width=0.47\textwidth]{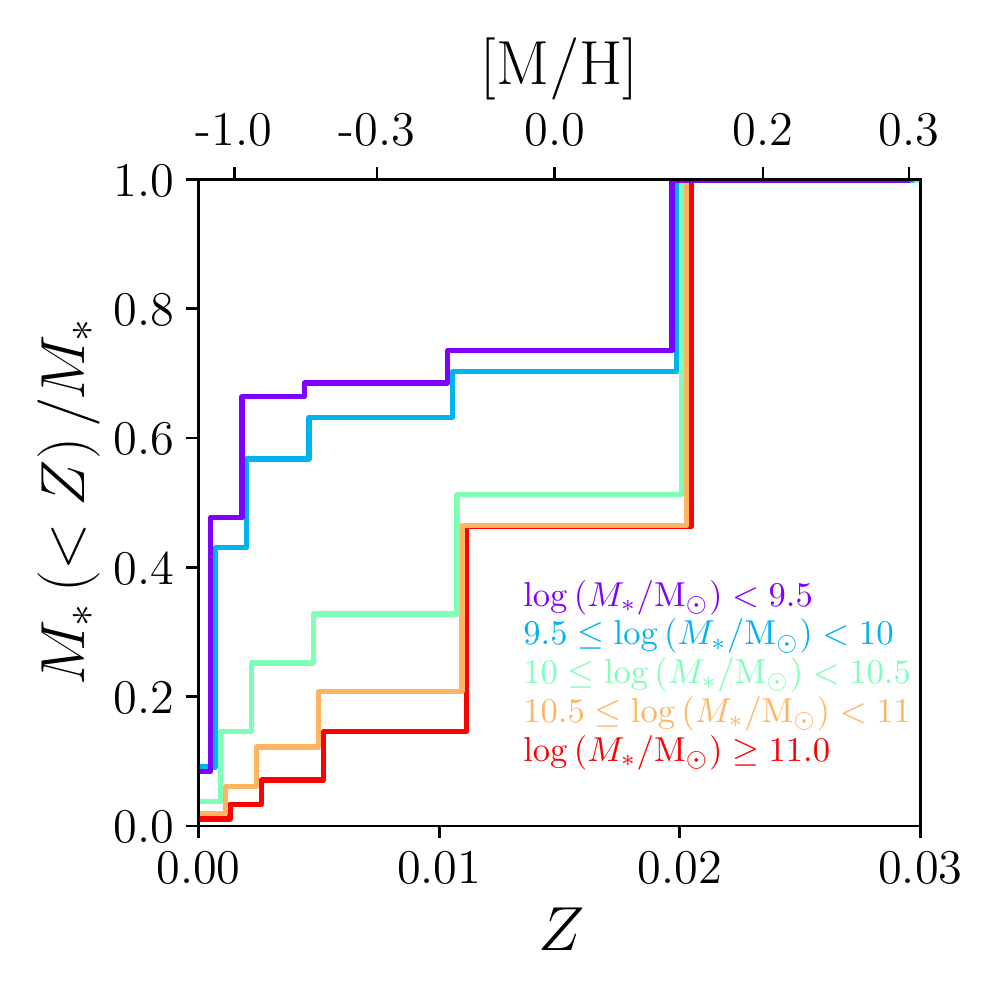}
    \caption{CMDFs for the spiral galaxies in the MaNGA sample, binned by stellar mass. The histograms show the median value for the CMDF within each mass bin, normalised by the total mass of each galaxy.}
    \label{fig:Median_MDF_total}
\end{figure}

The resulting CMDFs are presented in Figure~\ref{fig:Median_MDF_total}. In order to investigate any trend with galaxy mass, we have combined the galaxies into five logarithmically-spaced mass bins, normalised each galaxy by its total stellar mass, and calculated the median normalised CMDF within each bin. The step-like nature of the resulting cumulative functions reflects the relatively small number of template metallicities used in the fitting process, which, in turn, is determined by the limited amount of information that can be derived when decomposing such integrated spectral data.

It is immediately apparent from Figure~\ref{fig:Median_MDF_total} that the shape of a galaxy's CMDF depends strongly on stellar mass. Higher mass galaxies show a steepening CMDF, indicating a relative paucity of low-metallicity stars. Like their kin the Milky Way -- a galaxy of stellar mass ${\sim} 5 \times 10^{10} \ \rm M_{\odot}$ \citep{McMillan2017TheWay} -- they show a G-dwarf problem, which, comparison to Figure~\ref{fig:Theoretical_MDFs} confirms, is resolved if these systems are modelled as accreting boxes. By contrast, spiral galaxies with stellar masses of less than $10^{10} \ \rm M_{\odot}$ show a rapid initial rise in $M_{*}(< Z)$, reflecting their much greater proportion of low-metallicity stars, and matching rather well to the closed box model shown in Figure~\ref{fig:Theoretical_MDFs}. This finding builds on the significance of the much smaller sample studied by \citet{Mejia-Narvaez2020TheSurvey}, who found evidence that the distribution of metallicities is broader in lower mass spiral galaxies. It also fits with what has already been gleaned from the other axis in this population decomposition of MaNGA data, the time evolution of star formation, in which it was found that more massive spiral galaxies formed most of their stars in a relatively short period of time, whereas the less massive spiral systems have been slowly but steadily forming their stellar content over most of the lifetime of the Universe (Peterken et al.\ 2021, \emph{submitted}). It would appear that in the more massive galaxies, in order to keep up with the demand for gas to make more stars, largely unmixed pristine gas is pulled in to add to material enriched by the previous generations, making them produce a much larger fraction of high-metallicity stars in what is effectively an accreting box system. By contrast, the more leisurely star formation rate of the lower mass spirals affords them the opportunity to mix recycled gas thoroughly between stellar generations, making them behave as close to closed boxes. While the Milky Way is entirely typical of spiral galaxies of its size in displaying the G-dwarf problem caused by such systems' rush to make stars, lower-mass spiral galaxies avoid the issue by taking their time.

\section*{Data Availability}

This publication uses the team-internal MPL-9 MaNGA science data products. The full sample of data used here will be publicly released in 2021 as part of SDSS DR17.

\section*{Acknowledgements}

We thank the anonymous referee for their very positive comments and suggestions which have improved this manuscript.

This research was supported by funding from the Science and Technology Facilities Council (STFC).

Funding for the Sloan Digital Sky Survey IV has been provided by the Alfred P. Sloan Foundation, the U.S. Department of Energy Office of Science, and the Participating Institutions. SDSS acknowledges support and resources from the Center for High-Performance Computing at the University of Utah. The SDSS website is \url{www.sdss.org}.

SDSS is managed by the Astrophysical Research Consortium for the Participating Institutions of the SDSS Collaboration including the Brazilian Participation Group, the Carnegie Institution for Science, Carnegie Mellon University, the Chilean Participation Group, the French Participation Group, Harvard-Smithsonian Center for Astrophysics, Instituto de Astrof{\'i}sica de Canarias, The Johns Hopkins University, Kavli Institute for the Physics and Mathematics of the Universe (IPMU) / University of Tokyo, the Korean Participation Group, Lawrence Berkeley National Laboratory, Leibniz Institut f{\"u}r Astrophysik Potsdam (AIP), Max-Planck-Institut f{\"u}r Astronomie (MPIA Heidelberg), Max-Planck-Institut f{\"u}r Astrophysik (MPA Garching), Max-Planck-Institut f{\"u}r Extraterrestrische Physik (MPE), National Astronomical Observatories of China, New Mexico State University, New York University, University of Notre Dame, Observat{\'o}rio Nacional / MCTI, The Ohio State University, Pennsylvania State University, Shanghai Astronomical Observatory, United Kingdom Participation Group, Universidad Nacional Aut{\'o}noma de M{\'e}xico, University of Arizona, University of Colorado Boulder, University of Oxford, University of Portsmouth, University of Utah, University of Virginia, University of Washington, University of Wisconsin, Vanderbilt University, and Yale University.

This research made use of \texttt{Astropy} \citep{Robitaille2013Astropy:Astronomy}; \texttt{Marvin} \citep{Cherinka2018Marvin:Set}; \texttt{Matplotlib} \citep{Hunter2007Matplotlib:Environment}; \texttt{NumPy} \citep{VanderWalt2011TheComputation}; \texttt{SciPy} \citep{Virtanen2020SciPyPython}; and \texttt{TOPCAT} \citep{Taylor2005TOPCATSoftware}.

%%%%%%%%%%%%%%%%%%%%%%%%%%%%%%%%%%%%%%%%%%%%%%%%%%



\bibliographystyle{mnras}
\bibliography{main.bib} % if your bibtex file is called example.bib

%%%%%%%%%%%%%%%%%%%%%%%%%%%%%%%%%%%%%%%%%%%%%%%%%%

%%%%%%%%%%%%%%%%% APPENDICES %%%%%%%%%%%%%%%%%%%%%

%%%%%%%%%%%%%%%%%%%%%%%%%%%%%%%%%%%%%%%%%%%%%%%%%%

% Don't change these lines
\bsp	% typesetting comment
\label{lastpage}
\end{document}